\pdfoutput=1

\documentclass[12pt,a4paper]{article}
\usepackage{ifthen} 
\newboolean{pdflatex}
\setboolean{pdflatex}{true} 

\newboolean{articletitles}
\setboolean{articletitles}{true} 

\newboolean{uprightparticles}
\setboolean{uprightparticles}{false} 

\newboolean{inbibliography}
\setboolean{inbibliography}{false} 

\usepackage[top=1in, bottom=1.25in, left=1in, right=1in]{geometry}

\usepackage[normalem]{ulem}
\usepackage{microtype}
\usepackage{lineno}  
\usepackage{xspace} 
\usepackage{caption} 

\usepackage{graphicx}  
\usepackage{color}
\usepackage{colortbl}
\graphicspath{{./figs/}} 

\usepackage{amsmath} 
\usepackage{amssymb}
\usepackage{amsfonts}
\usepackage{upgreek} 

\newcommand*\patchAmsMathEnvironmentForLineno[1]{%
\expandafter\let\csname old#1\expandafter\endcsname\csname #1\endcsname
\expandafter\let\csname oldend#1\expandafter\endcsname\csname
end#1\endcsname
 \renewenvironment{#1}%
   {\linenomath\csname old#1\endcsname}%
   {\csname oldend#1\endcsname\endlinenomath}%
}
\newcommand*\patchBothAmsMathEnvironmentsForLineno[1]{%
  \patchAmsMathEnvironmentForLineno{#1}%
  \patchAmsMathEnvironmentForLineno{#1*}%
}
\AtBeginDocument{%
\patchBothAmsMathEnvironmentsForLineno{equation}%
\patchBothAmsMathEnvironmentsForLineno{align}%
\patchBothAmsMathEnvironmentsForLineno{flalign}%
\patchBothAmsMathEnvironmentsForLineno{alignat}%
\patchBothAmsMathEnvironmentsForLineno{gather}%
\patchBothAmsMathEnvironmentsForLineno{multline}%
\patchBothAmsMathEnvironmentsForLineno{eqnarray}%
}

\usepackage{hyperref}    
\usepackage[all]{hypcap} 

\usepackage{xspace} 
\usepackage{upgreek}

\def\lhcb {\mbox{LHCb}\xspace}

\def\MagUp {\mbox{\em Mag\kern -0.05em Up}\xspace}

\ifthenelse{\boolean{uprightparticles}}%
{

 \def\Peta        {\ensuremath{\upeta}\xspace}

 \def\Ppi         {\ensuremath{\uppi}\xspace}

 \def\Pphi        {\ensuremath{\upphi}\xspace}

 \def\PDelta      {\ensuremath{\Delta}\xspace}                 
 \def\PXi      {\ensuremath{\Xi}\xspace}                 
 \def\PLambda      {\ensuremath{\Lambda}\xspace}                 
 \def\PSigma      {\ensuremath{\Sigma}\xspace}                 
 \def\POmega      {\ensuremath{\Omega}\xspace}                 
 \def\PUpsilon      {\ensuremath{\Upsilon}\xspace}                 
 
 \def\PB      {\ensuremath{\mathrm{B}}\xspace}                 
                  
 \def\PD      {\ensuremath{\mathrm{D}}\xspace}

 \def\PK      {\ensuremath{\mathrm{K}}\xspace}

 \def\Pb      {\ensuremath{\mathrm{b}}\xspace}                 
 \def\Pc      {\ensuremath{\mathrm{c}}\xspace}                 
                  
 \def\Pe      {\ensuremath{\mathrm{e}}\xspace}

 \def\Pi      {\ensuremath{\mathrm{i}}\xspace}

 \def\Pp      {\ensuremath{\mathrm{p}}\xspace}

 \def\Ps      {\ensuremath{\mathrm{s}}\xspace}

}
{

 \def\Peta        {\ensuremath{\eta}\xspace}

 \def\Ppi         {\ensuremath{\pi}\xspace}

 \def\Pphi        {\ensuremath{\phi}\xspace}

 \mathchardef\PDelta="7101
 \mathchardef\PXi="7104
 \mathchardef\PLambda="7103
 \mathchardef\PSigma="7106
 \mathchardef\POmega="710A
 \mathchardef\PUpsilon="7107
                  
 \def\PB      {\ensuremath{B}\xspace}                 
                  
 \def\PD      {\ensuremath{D}\xspace}

 \def\PK      {\ensuremath{K}\xspace}

 \def\Pb      {\ensuremath{b}\xspace}                 
 \def\Pc      {\ensuremath{c}\xspace}                 
                  
 \def\Pe      {\ensuremath{e}\xspace}

 \def\Pi      {\ensuremath{i}\xspace}

 \def\Pp      {\ensuremath{p}\xspace}

 \def\Ps      {\ensuremath{s}\xspace}

}

\makeatletter
\ifcase \@ptsize \relax
  \newcommand{\miniscule}{\@setfontsize\miniscule{4}{5}}
\or
  \newcommand{\miniscule}{\@setfontsize\miniscule{5}{6}}
\or
  \newcommand{\miniscule}{\@setfontsize\miniscule{5}{6}}
\fi
\makeatother

\DeclareRobustCommand{\optbar}[1]{\shortstack{{\miniscule (\rule[.5ex]{1.25em}{.18mm})}
  \\ [-.7ex] $#1$}}

\def\epem       {{\ensuremath{\Pe^+\Pe^-}}\xspace}

\def\squark    {{\ensuremath{\Ps}}\xspace}
\def\squarkbar {{\ensuremath{\overline \squark}}\xspace}

\def\cquark    {{\ensuremath{\Pc}}\xspace}

\def\bquark    {{\ensuremath{\Pb}}\xspace}
\def\bquarkbar {{\ensuremath{\overline \bquark}}\xspace}

\def\pion   {{\ensuremath{\Ppi}}\xspace}
\def\piz    {{\ensuremath{\pion^0}}\xspace}

\def\pip    {{\ensuremath{\pion^+}}\xspace}
\def\pim    {{\ensuremath{\pion^-}}\xspace}

\def\kaon    {{\ensuremath{\PK}}\xspace}
  \def\Kbar    {{\kern 0.2em\overline{\kern -0.2em \PK}{}}\xspace}

\def\KorKbar    {\kern 0.18em\optbar{\kern -0.18em K}{}\xspace}

\def\Kp      {{\ensuremath{\kaon^+}}\xspace}
\def\Km      {{\ensuremath{\kaon^-}}\xspace}

\def\KS      {{\ensuremath{\kaon^0_{\mathrm{ \scriptscriptstyle S}}}}\xspace}

\def\Kstarz  {{\ensuremath{\kaon^{*0}}}\xspace}

\newcommand{\etapr}{\ensuremath{\Peta^{\prime}}\xspace}
\newcommand{\phiz}{\ensuremath{\Pphi}\xspace}

  \def\Dbar    {{\kern 0.2em\overline{\kern -0.2em \PD}{}}\xspace}
\def\D       {{\ensuremath{\PD}}\xspace}

\def\DorDbar    {\kern 0.18em\optbar{\kern -0.18em D}{}\xspace}
\def\Dz      {{\ensuremath{\D^0}}\xspace}

\def\Dstarp  {{\ensuremath{\D^{*+}}}\xspace}

\def\B       {{\ensuremath{\PB}}\xspace}
\def\Bbar    {{\ensuremath{\kern 0.18em\overline{\kern -0.18em \PB}{}}}\xspace}

\def\BorBbar    {\kern 0.18em\optbar{\kern -0.18em B}{}\xspace}
\def\Bz      {{\ensuremath{\B^0}}\xspace}

\def\Bu      {{\ensuremath{\B^+}}\xspace}

\def\Bp      {{\ensuremath{\Bu}}\xspace}

\def\Bd      {{\ensuremath{\B^0}}\xspace}
\def\Bs      {{\ensuremath{\B^0_\squark}}\xspace}

  \def\Y#1S{\ensuremath{\PUpsilon{(#1S)}}\xspace}

\def\proton      {{\ensuremath{\Pp}}\xspace}

\def\Lbar        {{\ensuremath{\kern 0.1em\overline{\kern -0.1em\PLambda}}}\xspace}
\def\LorLbar    {\kern 0.18em\optbar{\kern -0.18em \PLambda}{}\xspace}

\def\BF         {{\ensuremath{\mathcal{B}}}\xspace}

\def\BR         {\BF}
\newcommand{\decay}[2]{\ensuremath{#1\!\to #2}\xspace}         

\def\to                 {\ensuremath{\rightarrow}\xspace}

\def\CP                {{\ensuremath{C\!P}}\xspace}

\def\AT#1     {\ensuremath{A_{\mathrm{T}}^{#1}}\xspace}           

\def\C#1      {\ensuremath{\mathcal{C}_{#1}}\xspace}                       
\def\Cp#1     {\ensuremath{\mathcal{C}_{#1}^{'}}\xspace}                    
\def\Ceff#1   {\ensuremath{\mathcal{C}_{#1}^{\mathrm{(eff)}}}\xspace}        
\def\Cpeff#1  {\ensuremath{\mathcal{C}_{#1}^{'\mathrm{(eff)}}}\xspace}       
\def\Ope#1    {\ensuremath{\mathcal{O}_{#1}}\xspace}                       
\def\Opep#1   {\ensuremath{\mathcal{O}_{#1}^{'}}\xspace}                    

\newcommand{\tev}{\ifthenelse{\boolean{inbibliography}}{\ensuremath{~T\kern -0.05em eV}\xspace}{\ensuremath{\mathrm{\,Te\kern -0.1em V}}}\xspace}
\newcommand{\gev}{\ensuremath{\mathrm{\,Ge\kern -0.1em V}}\xspace}
\newcommand{\mev}{\ensuremath{\mathrm{\,Me\kern -0.1em V}}\xspace}
\newcommand{\kev}{\ensuremath{\mathrm{\,ke\kern -0.1em V}}\xspace}
\newcommand{\ev}{\ensuremath{\mathrm{\,e\kern -0.1em V}}\xspace}
\newcommand{\gevc}{\ensuremath{{\mathrm{\,Ge\kern -0.1em V\!/}c}}\xspace}
\newcommand{\mevc}{\ensuremath{{\mathrm{\,Me\kern -0.1em V\!/}c}}\xspace}
\newcommand{\gevcc}{\ensuremath{{\mathrm{\,Ge\kern -0.1em V\!/}c^2}}\xspace}
\newcommand{\gevgevcccc}{\ensuremath{{\mathrm{\,Ge\kern -0.1em V^2\!/}c^4}}\xspace}
\newcommand{\mevcc}{\ensuremath{{\mathrm{\,Me\kern -0.1em V\!/}c^2}}\xspace}

\def\mum  {\ensuremath{{\,\upmu\mathrm{m}}}\xspace}

\def\invfb   {\ensuremath{\mbox{\,fb}^{-1}}\xspace}

\newcommand{\chisq}{\ensuremath{\chi^2}\xspace}

\newcommand{\chisqip}{\ensuremath{\chi^2_{\text{IP}}}\xspace}

\def\gsim{{~\raise.15em\hbox{$>$}\kern-.85em
          \lower.35em\hbox{$\sim$}~}\xspace}
\def\lsim{{~\raise.15em\hbox{$<$}\kern-.85em
          \lower.35em\hbox{$\sim$}~}\xspace}

\def\sPlot{\mbox{\em sPlot}\xspace}

\def\pt         {\mbox{$p_{\mathrm{ T}}$}\xspace}

\def\evtgen     {\mbox{\textsc{EvtGen}}\xspace}

\def\geant      {\mbox{\textsc{Geant4}}\xspace}

\def\photos     {\mbox{\textsc{Photos}}\xspace}

\def\pythia     {\mbox{\textsc{Pythia}}\xspace}

\def\tell1  {TELL1\xspace}
\def\ukl1   {UKL1\xspace}

\usepackage{cite} 
\usepackage{mciteplus}

\usepackage{longtable} 

\begin{document}

\renewcommand{\thefootnote}{\fnsymbol{footnote}}
\setcounter{footnote}{1}


\begin{titlepage}
\pagenumbering{roman}

\vspace*{-1.5cm}
\centerline{\large EUROPEAN ORGANIZATION FOR NUCLEAR RESEARCH (CERN)}
\vspace*{1.5cm}
\noindent
\begin{tabular*}{\linewidth}{lc@{\extracolsep{\fill}}r@{\extracolsep{0pt}}}
\ifthenelse{\boolean{pdflatex}}
{\vspace*{-2.7cm}\mbox{\!\!\!\includegraphics[width=.14\textwidth]{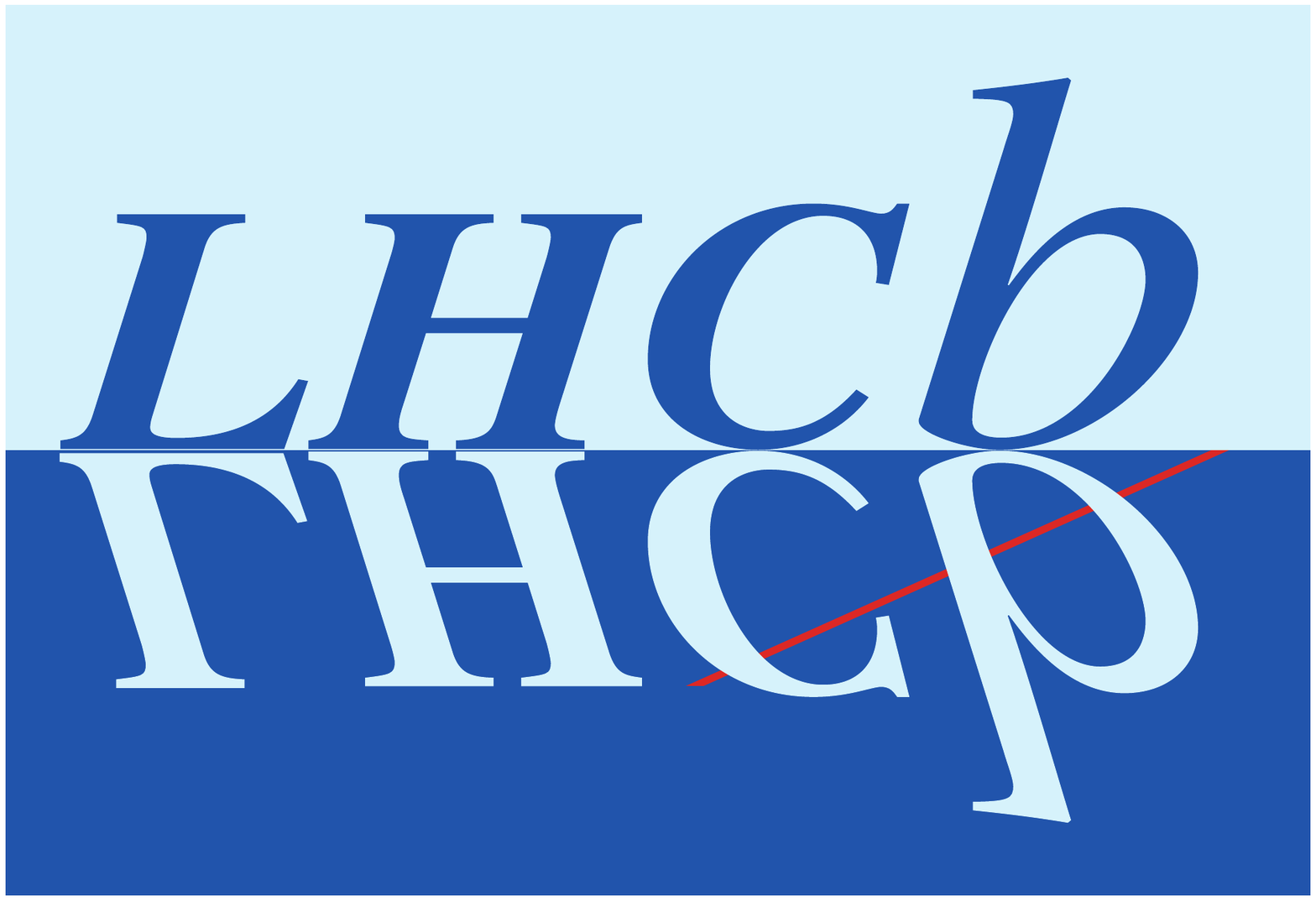}} & &}%
{\vspace*{-1.2cm}\mbox{\!\!\!\includegraphics[width=.12\textwidth]{lhcb-logo.eps}} & &}%
\\
 & & CERN-EP-2016-304 \\  
 & & LHCb-PAPER-2016-060 \\  
 & & 21 March 2017 \\ 
 & & \\
\end{tabular*}

\vspace*{4.0cm}

{\normalfont\bfseries\boldmath\huge
\begin{center}
  Search for the \decay{\Bs}{\etapr\phiz} decay
\end{center}
}

\vspace*{2.0cm}

\begin{center}
The LHCb collaboration\footnote{Authors are listed at the end of this paper.}
\end{center}

\vspace{\fill}

\begin{abstract}
  \noindent
\noindent
A search for the charmless \decay{\Bs}{\etapr\phiz} decay is performed
using $pp$ collision data collected by the LHCb experiment at centre-of-mass energies of $7$ and~$8\tev$, corresponding to an integrated luminosity of 3\invfb.
No signal is observed and
upper limits on the \decay{\Bs}{\etapr\phiz} branching fraction are set to
\mbox{$0.82\times 10^{-6}$} at $90$\%
and \mbox{$1.01\times 10^{-6}$} at $95$\% confidence level.
  
\end{abstract}

\vspace*{2.0cm}

\begin{center}
Published in JHEP 05 (2017) 158
\end{center}

\vspace{\fill}

{\footnotesize 
\centerline{\copyright~CERN on behalf of the \lhcb collaboration, licence \href{http://creativecommons.org/licenses/by/4.0/}{CC-BY-4.0}.}}
\vspace*{2mm}

\end{titlepage}


\newpage
\setcounter{page}{2}
\mbox{~}
%
%
%
%

\cleardoublepage


\renewcommand{\thefootnote}{\arabic{footnote}}
\setcounter{footnote}{0}


\pagestyle{plain} 
\setcounter{page}{1}
\pagenumbering{arabic}


%

\newcommand{\proposal}[2]{{\color{red}\sout{#1}#2}}
\newcommand{\proposaltwo}[1]{{\color{blue}#1}}
\section{Introduction}
\label{sec:Introduction}

Charmless hadronic decays of beauty hadrons proceed predominantly through tree-level $b\to u$ and
loop-level (penguin) $b\to s$ weak transitions. In the Standard Model the amplitudes of these processes, suppressed
compared to the dominant tree $b\to c$ transition governing charmed decays, usually have similar magnitudes and give rise to possibly large
violation of the charge-parity (\CP) symmetry.
Therefore, charmless decays of $B$ mesons should be sensitive to additional amplitudes from new, heavy particles,
contributing to the loop-level transitions~\cite{Zhang:2000ic}. 

Charmless hadronic \Bp and \Bz decays
\footnote{The inclusion of charge-conjugate processes is implied throughout.}
have been the subject of extensive studies, both experimentally,
at hadron and \epem colliders, and theoretically.
The phenomenological understanding that has emerged allows
predictions to be made for charmless \Bs decays, as will be illustrated in the following. 
In the ongoing effort to test these predictions experimentally,
the LHCb experiment has 
recently observed the
decay\footnote{The notations $\etapr$ and $\phiz$ refer to the $\etapr(958)$ and 
$\phi(1020)$ mesons.}
\decay{\Bs}{\etapr\etapr}. The relatively large measured branching fraction
$\BF(\decay{\Bs}{\etapr\etapr}) = (33.1 \pm 7.1)\times10^{-6}$
is consistent with 
Standard Model
expectations~\cite{LHCb-PAPER-2014-065}.
However, the knowledge about charmless hadronic \Bs decays
into light pseudoscalar~(P) and vector~(V) mesons
is still limited. Further measurements will help to better
constrain phenomenological models, the uncertainties of which
often translate into a major contribution to the theoretical
uncertainties in searches for physics beyond the Standard Model.

\begin{figure}[t]
\begin{center}
\includegraphics[width=0.5\textwidth]{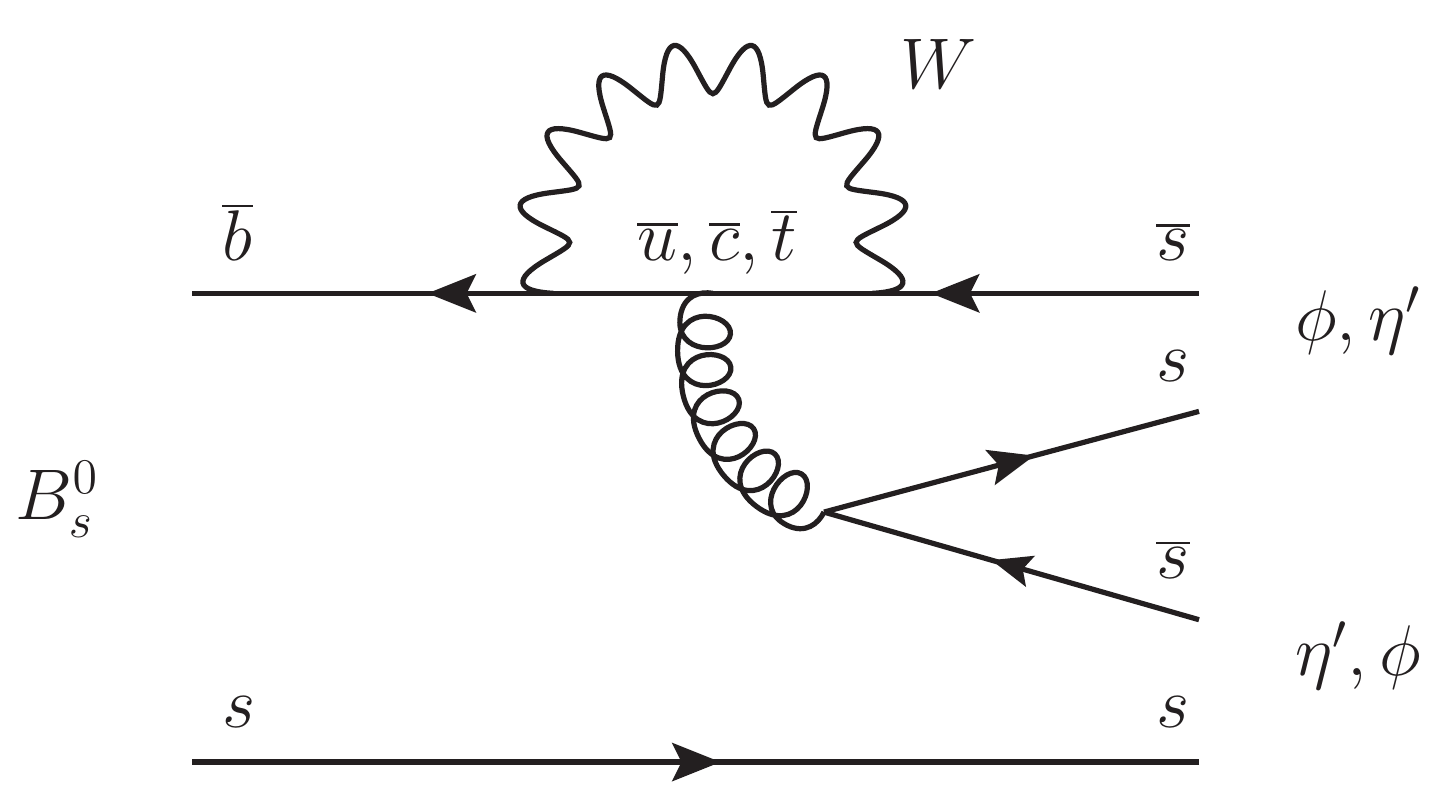}
\end{center}
\caption{Lowest-order diagrams for the \decay{\Bs}{\etapr\phiz} decay.
The spectator quark can become part of either the $\etapr$ or the $\phiz$ meson, forming two different amplitudes (called PV and VP in the text).}
\label{fig:feynmandiagram}
\end{figure}

The decay \decay{\Bs}{\etapr\phiz} proceeds predominantly through $\bquarkbar\to\squarkbar\squark\squarkbar$
transitions, as illustrated in Fig.~\ref{fig:feynmandiagram}. 
It is of particular interest in constraining phenomenological models, as predictions
for its branching fraction cover a wide range, typically from $0.1\times 10^{-6}$ to $20\times 10^{-6}$, 
with large uncertainties that reflect the limited knowledge of form factors,
penguin contributions, the $\omega-\phiz$ mixing angle, or the \squark-quark mass. 
The decay \decay{\Bs}{\etapr\phiz} has been studied in the framework of
QCD factorisation~\cite{Beneke:2003zv, Cheng:2009mu},
perturbative QCD~\cite{Ali:2007ff, Chen:2007qm},
soft-collinear effective theory (SCET)~\cite{Wang:2008rk},
SU(3) flavour symmetry~\cite{Cheng:2014rfa},
and factorisation-assisted topological (FAT) amplitude approach~\cite{Zhou:2016jkv}.
Table~\ref{tab:predictions} presents the available predictions for $\BF(\decay{\Bs}{\etapr\phiz})$.

In QCD factorisation, predictions for $\BF(\decay{\Bs}{\etapr\phiz})$ are generally small
because the spectator quark can become part of either the $\etapr$ or the $\phiz$ meson (see Fig.~\ref{fig:feynmandiagram}), leading to
a strong cancellation between the PV and VP amplitudes contributing to the $\etapr\phiz$ final state~\cite{Beneke:2003zv}.
Such cancellation does not occur in the symmetric \decay{\Bs}{\etapr\etapr} (PP) and \decay{\Bs}{\phiz\phiz} (VV) decays.
However, other values of the form factor for the $\Bs$ to $\phiz$ transitions can lead to enhancements of the branching fraction by
more than an order of magnitude~\cite{Cheng:2009mu}.
The measurement of $\BF(\decay{\Bs}{\etapr\phiz})$ is therefore important to improve
the knowledge of the $\Bs$ to $\phiz$ form factor and the accuracy of model predictions.

The comparison of QCD factorisation~\cite{Beneke:2003zv, Cheng:2009mu},
perturbative QCD~\cite{Ali:2007ff},
and SCET~\cite{Wang:2008rk} calculations shows that the hierarchy of branching
fractions in \decay{\Bs}{\etapr\phiz} and \decay{\Bs}{\eta\phiz} decays is sensitive to the size of the colour-suppressed
QCD penguin loop, which is estimated to be large in perturbative QCD~\cite{Ali:2007ff},
and to ``gluonic charming penguins'', which play an important role in SCET calculations~\cite{Wang:2008rk}.
Future measurements of both decay modes will provide useful information on these loop contributions.

This paper presents a search for the \decay{\Bs}{\etapr\phiz} decay with the \lhcb detector.
The results are based on a data sample collected during the 2011 and 2012 $\proton\proton$ collision
runs of the Large Hadron Collider at centre-of-mass energies of $7$ and $8~\tev$, respectively,
corresponding to a total integrated luminosity of $3\invfb$.

The signal \decay{\Bs}{\etapr\phiz} and normalisation \decay{\Bp}{\etapr\Kp}  candidates
are reconstructed through the decays \decay{\etapr}{\pip\pim\gamma} and \decay{\phiz}{\Kp\Km}.
The $\Bs \to \etapr\phiz$ branching fraction is determined with respect to the \decay{\Bp}{\etapr\Kp}  mode according to
\begin{equation}
\BF(\Bs \to \etapr \phiz) = \frac{\BF(B^+ \to \etapr K^+)}{\BF(\phiz\to K^+K^-)}\times \frac{f_u}{f_s} 
\times  \frac{N(\Bs \to \etapr \phiz)}{N(B^+ \to \etapr K^+)}
\times  \frac{\epsilon(B^+ \to \etapr K^+)}{\epsilon(\Bs \to \etapr \phiz)} \,,
\label{eq:master}
\end{equation}
where \mbox{$\BF(B^+ \to \etapr K^+) = (70.6\pm 2.5)\times 10^{-6}$}~\cite{PDG2016}, 
$\BF(\phiz\to K^+K^-) = 0.489\pm 0.005$~\cite{PDG2016}, 
$f_u/f_s$ is the $B^+/\Bs$ production ratio assumed to be equal to the  $B^0/\Bs$ production ratio \mbox{$f_d/f_s = 1/(0.259 \pm 0.015)$}~\cite{fsfd},
and $\epsilon(\Bs \to \etapr \phiz)$ and $\epsilon(B^+ \to \etapr K^+)$ are the total efficiencies of the signal and normalisation modes, respectively.
The ratio of the observed yields  $N(\Bs \to \etapr \phiz)/N(B^+ \to \etapr K^+)$ is obtained from a two-dimensional fit to the invariant mass
distributions of the $\etapr$ and the $B$ candidates, performed simultaneously on the signal and normalisation modes.

\begin{table}[t]
\caption{\small Theoretical predictions for the \decay{\Bs}{\etapr\phiz} branching fraction.}
\label{tab:predictions}
\renewcommand{\arraystretch}{1.2} 
\begin{center}\begin{tabular}{lccc}
\hline
Theory approach &  $ \BF\ (10^{-6})$ & Reference  \\ 
\hline
QCD factorisation &  $0.05 ^{+1.18}_{-0.19}$ & \cite{Beneke:2003zv}  \\ 
QCD factorisation &   $\,2.2  ^{+9.4}_{-3.1}$ & \cite{Cheng:2009mu}  \\
Perturbative QCD  & $0.19 ^{+0.20}_{-0.13}$ & \cite{Ali:2007ff}\\
Perturbative QCD &  $20.0 ^{+16.3}_{-9.1}$ & \cite{Chen:2007qm} \\
SCET &   $\,4.3^{+5.2}_{-3.6}$ & \cite{Wang:2008rk} \\
SU(3) flavour symmetry  &  $\:\;5.5 \pm 1.8$ & \cite{Cheng:2014rfa} \\
FAT & $13.0 \pm 1.6$ & \cite{Zhou:2016jkv} \\
\hline
\end{tabular}\end{center} 
\end{table}

\section{Detector and simulation}
\label{sec:Detector}

The \lhcb detector~\cite{Alves:2008zz,LHCb-DP-2014-002} is a single-arm forward
spectrometer covering the \mbox{pseudorapidity} range $2<\eta <5$,
designed for the study of particles containing \bquark or \cquark
quarks. The detector includes a high-precision tracking system
consisting of a silicon-strip vertex detector surrounding the $pp$
interaction region, a large-area silicon-strip detector located
upstream of a dipole magnet with a bending power of about
$4{\mathrm{\,Tm}}$, and three stations of silicon-strip detectors and straw
drift tubes placed downstream of the magnet.
The tracking system provides a measurement of momentum of charged particles with
a relative uncertainty that varies from 0.5\% at low momentum to 1.0\% at 200\gevc.
The minimum distance of a track to a $pp$-collision point (primary vertex), the impact parameter, 
is measured with a resolution of $(15+29/\pt)\mum$,
where \pt is the component of the momentum transverse to the beam, in\,\gevc.
Different types of charged hadrons are distinguished using information
from two ring-imaging Cherenkov detectors. 
Photons, electrons and hadrons are identified by a calorimeter system consisting of
scintillating-pad (SPD) and preshower detectors, an electromagnetic
calorimeter and a hadronic calorimeter. Muons are identified by a
system composed of alternating layers of iron and multiwire
proportional chambers.

The trigger~\cite{LHCb-DP-2012-004} consists of a
hardware stage, based on information from the calorimeter and muon
systems, followed by a software stage, which applies a  full event reconstruction.
The $B$ decays of interest are triggered at the hardware stage, either by one of the decay products
depositing a transverse energy greater than 3.5\gev in the hadron calorimeter, or by other high-\pt particles produced in the $pp$~collision.
The software trigger requires a two-, three- or four-track
secondary vertex with a significant displacement from the primary vertices.
At least one charged particle
must have a transverse momentum $\pt > 1.7\gevc$ and be
inconsistent with originating from a primary vertex.
A multivariate algorithm~\cite{BBDT} is used for
the identification of secondary vertices consistent with the decay
of a \bquark hadron.

Simulated decays are used to optimise the event selection
and to evaluate the selection efficiencies.
In the simulation, $pp$ collisions are generated using \pythia8~\cite{Sjostrand:2006za,*Sjostrand:2007gs} 
with a specific \lhcb
configuration~\cite{LHCb-PROC-2010-056}.  Decays of hadronic particles
are described by \evtgen~\cite{Lange:2001uf}, in which final-state
radiation is generated using \photos~\cite{Golonka:2005pn}. The
interaction of the generated particles with the detector, and its response,
are implemented using the \geant
toolkit~\cite{Allison:2006ve, *Agostinelli:2002hh} as described in
Ref.~\cite{LHCb-PROC-2011-006}.

\section{Event selection}

The selection of the signal \decay{\Bs}{\etapr\phiz} and normalisation \decay{\Bp}{\etapr\Kp}
candidates, generically referred to as B candidates, is optimised for
the signal. Wherever possible, the same selection criteria are applied
for the normalisation channel.

Only good-quality tracks identified  
as pions or kaons~\cite{LHCb-DP-2014-002} and inconsistent with originating from any primary vertex 
are used. 
Tracks used to reconstruct an $\etapr$ or $\phiz$ candidate are each required to be consistent 
with coming from a common secondary vertex and to have $\pt > 0.4\gevc$. 
The $\pip\pim$ invariant mass in the $\etapr$ decay must be larger than $0.52\gevcc$ to reject \decay{\KS}{\pip\pim} decays.
Photon candidates must be of good quality~\cite{LHCb-DP-2014-002} and have $\pt> 0.3\gevc$.
The invariant masses of the $\etapr$ and $\phiz$ candidates must  satisfy
\mbox{$0.88<m_{\pi\pi\gamma}<1.04\gevcc$} and \mbox{$1.005<m_{KK}<1.035\gevcc$}. 
An $\etapr$ candidate is combined with a candidate $\phiz$ meson (or a charged kaon with $\pt> 1\gevc$) to make a $\Bs$ (or $\Bp$) candidate.
Each $B$ candidate is required to have a good-quality vertex, by imposing a loose requirement  of the $\chi^{2}$ of the vertex fit ($\chi^2 < 6$), and $\pt>1.5\gevc$. 
The invariant masses of the \Bs and \Bp candidates, 
computed after constraining the $\pi^+\pi^-\gamma$ mass to the nominal $\etapr$ mass~\cite{PDG2016},
are required to satisfy
\mbox{$5.0<m_{\etapr KK}<5.6\gevcc$} and \mbox{$5.0<m_{\etapr K}<5.5\gevcc$}, respectively.


To  further 
separate signal from background, boosted decision trees~(BDTs) based on the AdaBoost algorithm~\cite{Breiman,AdaBoost} are used.
Different BDTs are used for the signal and normalisation channels.
Each BDT is trained, tested and optimised on fully simulated signal decays and background taken from data.
The background  consist of events in the mass range \mbox{$5.0<m_{\etapr KK}<5.6\gevcc$} (\mbox{$5.0<m_{\etapr K}<5.5\gevcc$}) excluding the signal region defined below.

To minimise statistical and systematic uncertainties, the BDT algorithm uses input variables
that provide significant background rejection, are well modelled in simulation, and are defined
for both the signal and normalisation channels. 
Nine variables are used as input to each BDT.
Two variables are related to the kinematics of the final-state particles: the transverse momenta of the photon and the $\etapr$ meson.
Three variables describe the topology of the $\B$ candidate:
the $B$-candidate flight distance,
the cosine of the angle between the reconstructed $\B$~momentum and the vector pointing from the associated primary vertex to the $\B$ decay vertex,
and the impact parameter of the $\B$ candidate with respect to its associated primary vertex.
The associated primary vertex is the primary vertex with respect to which the $B$ candidate has the smallest $\chisqip$,
where $\chisqip$ is defined as the difference in the vertex-fit $\chi^2$ of the selected primary vertex reconstructed with or without the considered particle.
Three variables are related to the $\B$-candidate vertex:
the vertex-fit quality, characterised by its $\chi^2$,
and two vertex isolation variables 
defined as the smallest vertex-fit \chisq\ values obtained when adding to the vertex in turn either all single tracks
or all pairs of tracks from the set of tracks that are not assigned to the $B$ candidate.
The last variable is the sum of the $\chisqip$ of the charged particles used to form the $\B$ candidate, calculated with respect to the associated primary vertex.
The photon $\pt$ and the $B$-candidate impact parameter provide the best background discrimination.
The BDT is trained for the full data set, irrespective of the $pp$ collision energy. 
To minimise biases in the final selection, both the data and simulated samples are randomly divided
into two subsamples and two BDTs are defined. Each  BDT is trained, tested and optimised
on one subsample, and then applied to the other subsample for the candidate selection~\cite{Blum:1999:BHB:307400.307439}.
The selected candidates from both subsamples are then merged into a single sample for the next stage of the analysis.

The requirement on the BDT output is chosen to maximise the
figure of merit \mbox{$\epsilon(\decay{\Bs}{\etapr\phiz})/(a/2+\sqrt{N_B})$}~\cite{Punzi:2003bu},
where $a = 5$  is the target signal significance,
and $N_B$ is the number of background events in the signal region estimated from the \Bs mass sidebands.
The signal region is defined as the \Bs mass range $5.287-5.446\gevcc$, corresponding approximately to $7$ times the mass resolution.
The optimised BDT requirement has an efficiency of
$59\%$ 
for \decay{\Bs}{\etapr\phiz} decays, while rejecting 
$93\%$ 
of the combinatorial background in the signal region. 
As a check, an alternative optimisation is performed: 
for various values of the \decay{\Bs}{\etapr\phiz} branching fraction, pseudoexperiments are generated
with a model containing only signal and combinatorial background,
and then are analysed with a simple two-dimensional maximum likelihood fit to the \Bs and $\etapr$ masses.
The signal significance, determined using Wilks' theorem~\cite{Wilks:1938dza}, is found to reach its maximum 
for a BDT requirement in agreement with that obtained using the method of Ref.~\cite{Punzi:2003bu}.

In events containing multiple candidates ($\lesssim 3\%$), the candidate with the best identified photon is kept. 
The full selection described above retains $430$~\decay{\Bs}{\etapr \phiz} candidates and $22\,681$~\decay{\Bp}{\etapr\Kp} candidates for further analysis. 



\begin{table}[!t]
 \caption{Relative systematic uncertainties on the efficiency ratio $\epsilon(\decay{\Bp}{\etapr\Kp}) / \epsilon(\decay{\Bs}{\etapr \phiz})$. 
}
\begin{center}
\begin{tabular}{lc}
\hline
Source & Relative uncertainty [\%]  \\
\hline
BDT efficiency calibration  & 2.5 \\ 
PID efficiency calibration & 1.1 \\ 
Trigger efficiency calibration & 2.3 \\ 
SPD multiplicity (mismodelling) & 0.9 \\ 
Track reconstruction & 0.4 \\ 
Photon reconstruction & 0.1 \\ 
Hadronic interactions  & 1.4 \\
Simulation statistics & 1.6 \\ 
\hline
Total   & 4.3 \\ 
\hline
\end{tabular}\end{center}
\label{tab:Systematics_eff}
\end{table}

Selection efficiencies are evaluated with simulated data, except those of the particle identification (PID) requirements
and the hardware trigger, for which  calibration data are used. 
Systematic uncertainties on the efficiency ratio
\mbox{$\epsilon(\decay{\Bp}{\etapr\Kp}) / \epsilon(\decay{\Bs}{\etapr \phi})$}
are summarised in Table~\ref{tab:Systematics_eff}.
The BDT algorithms are validated using the normalisation channel as proxy for the signal, and
by comparing the distributions obtained with the \sPlot technique~\cite{Pivk:2004ty} of the 
nine input variables and the BDT output variable.
The difference between the efficiencies in data and simulation of the BDT requirement for the normalisation channel
is used as a measure of the systematic uncertainty on the BDT efficiency. The correlation evaluated in simulation
between the BDT variables for \decay{\Bs}{\etapr \phi} and \decay{\Bp}{\etapr\Kp} is then used to determine
the systematic uncertainty on the ratio of the BDT efficiencies.

Another systematic effect on the determination of the efficiency ratio is the uncertainty on the PID efficiency,
which is determined as a function of kinematic parameters using a clean 
high-statistics sample of kaons and pions from \mbox{$\decay{\Dstarp}{\Dz (\to \Km\pip)\pip}$} decays~\cite{LHCb-DP-2012-003}.
The uncertainty on the trigger efficiency,
which is mostly due to the computation of the hardware-stage trigger efficiency,
is estimated
with simulated data 
by varying the value of the minimum transverse energy requirement used in the trigger decision.
An uncertainty is assigned on the efficiency ratio to take into account the mismodelling of the hit
multiplicity in the SPD, which is used as a discriminant variable at the
hardware stage of the trigger. This uncertainty is evaluated in simulation by varying the requirement on 
the SPD hit multiplicity.
Corrections determined from control channels are applied to the tracking and photon reconstruction efficiencies to account for
mismodelling effects in the simulation. The uncertainties on these corrections are quoted
as systematic uncertainties.
Since the correction to the tracking efficiency is obtained using muons, an additional uncertainty is needed to account for
hadronic interactions in the detector material~\cite{LHCb-DP-2013-002}. 
Finally, the limited statistics of the simulated  samples used in the evaluation of the efficiencies
is added as a source of uncertainty.
Combining all uncertainties in quadrature, the ratio of the selection efficiencies is 
\begin{equation}
\frac{\epsilon(\decay{\Bp}{\etapr\Kp})}{\epsilon(\decay{\Bs}{\etapr \phi})} = 1.83 \pm  0.08 \,.
\label{eq:eff_ratio}
\end{equation}

The selection requirements efficiently reject physics backgrounds such as \decay{\Bd}{\phiz\Kstarz} and modes
with resonances decaying strongly to $\Kp\pim\piz$, but not \decay{\Bs}{\phiz\phiz} decays
with one of the two $\phiz$ resonances decaying to $\pip\pim\piz$
and one of the photons from the \decay{\piz}{\gamma\gamma} decay not being reconstructed. 
From simulation studies and known branching fractions~\cite{PDG2016}, 
the number of \decay{\Bs}{\phiz\phiz} decays passing the selection is expected to be 
$104 \pm 34$.
Hence this background is included as a specific component in the  mass fit described below. 


\section{Mass fit}

The \decay{\Bs}{\etapr\phiz} signal yield is determined from a two-dimensional extended unbinned
maximum likelihood fit, where the signal is fitted simultaneously  with the normalisation channel \decay{\Bp}{\etapr \Kp}.
The observables used in the fit are the invariant masses $m_{\pi\pi\gamma}$ and $m_{\etapr KK}$ ($m_{\etapr K}$) 
for the sample of \decay{\Bs}{\etapr\phiz} (\decay{\Bp}{\etapr \Kp}) candidates. 

The sample of \decay{\Bs}{\etapr\phiz} candidates is described with a four-component model: 
the signal, the two combinatorial backgrounds with and without a true $\etapr$ resonance, 
and the \decay{\Bs}{\phiz\phiz} physics background, where one of the two $\phiz$ resonances decays to  the $\pip\pim\piz$ final state.
The sample of \decay{\Bp}{\etapr \Kp} candidates is modelled using three components: 
the signal and the two combinatorial backgrounds with and without a true $\etapr$ resonance. 
The yields of all components are free to vary in the fit.
The peaking components in the \Bs, \Bp and $\etapr$ mass spectra are described using
Gaussian functions modified with an exponential tail on each side.
While all  the tail parameters are fixed from simulation, the mean and the widths of the Gaussian functions
are free to vary in the fit,  
but the ratio of the widths of the peaking components in $m_{\etapr KK}$ and $m_{\etapr K}$ is fixed to the value obtained
in simulation and the difference between the \Bs and  \Bp masses is constrained to the known value~\cite{PDG2016}.
The $\etapr$ resonances in the two samples are modelled using a common function, with mean and width free in the fit.
The combinatorial components are described with linear functions, with the exception of the random combinations
in $m_{\etapr K}$, where a parabolic function is used.
To account for correlations between $m_{\etapr KK}$ and $m_{\pi\pi\gamma}$,
the \decay{\Bs}{\phiz\phiz} component is described with
a superposition of two-dimensional  Gaussian kernel functions~\cite{Cranmer:2000du} determined from simulation.
For all other components, in particular the signal, the correlation is negligible due to the \etapr mass constraint applied
in the computation of the $B$-candidate mass.
The fit procedure is validated on simulated samples containing the expected proportion of
background and signal events, according to various assumptions on $\BF(\Bs \to \etapr \phiz)$.
In particular, for $\BR(\decay{\Bs}{\etapr\phiz}) =4\times 10^{-6}$, a statistical significance corresponding
to more than $5$~standard deviations is observed in $74\%$ of the pseudoexperiments.

\begin{figure}[t]
\includegraphics[width=0.51\textwidth]{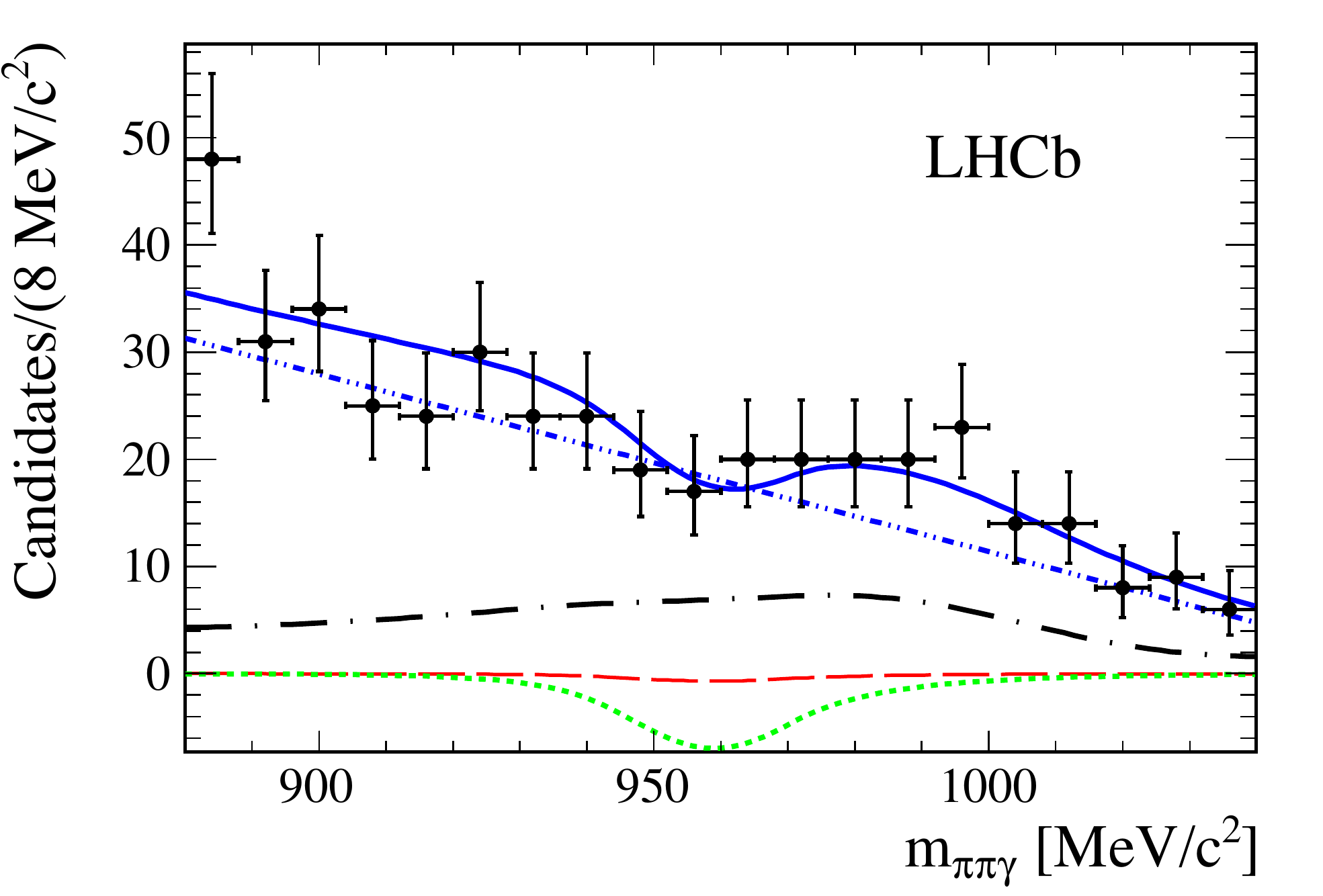}
\hspace{-1em}
\includegraphics[width=0.51\textwidth]{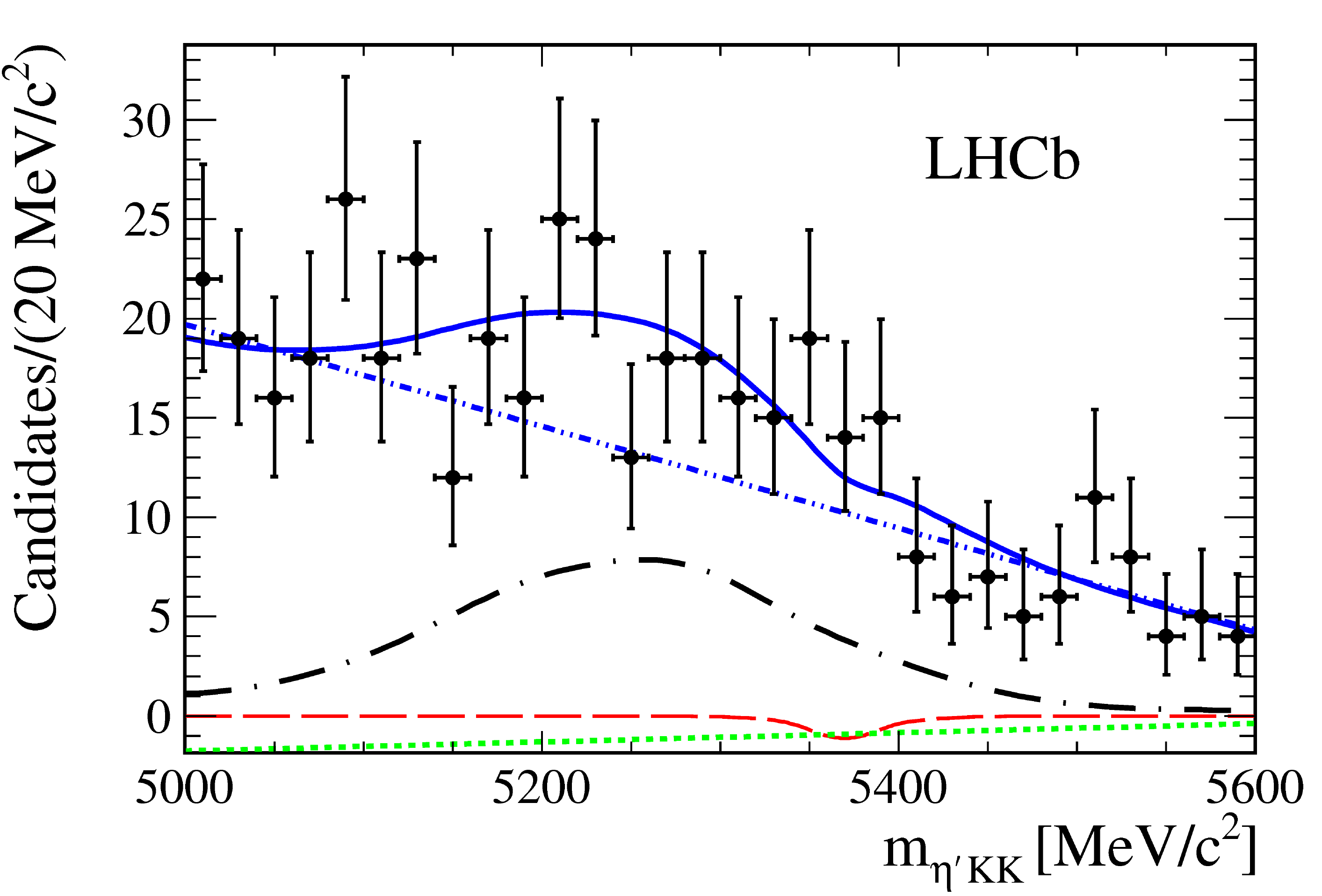}

\includegraphics[width=0.51\textwidth]{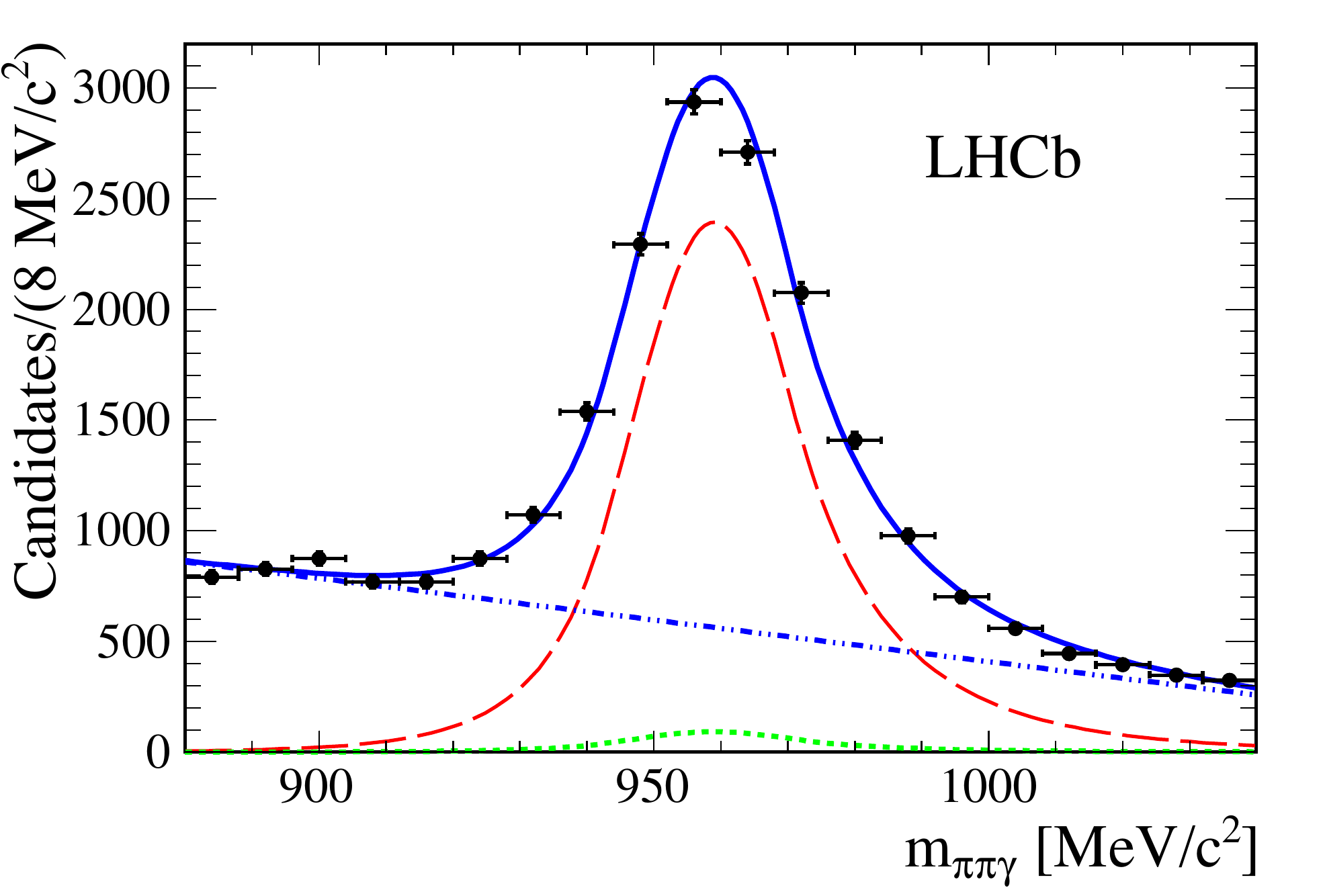}
\hspace{-1em}
\includegraphics[width=0.51\textwidth]{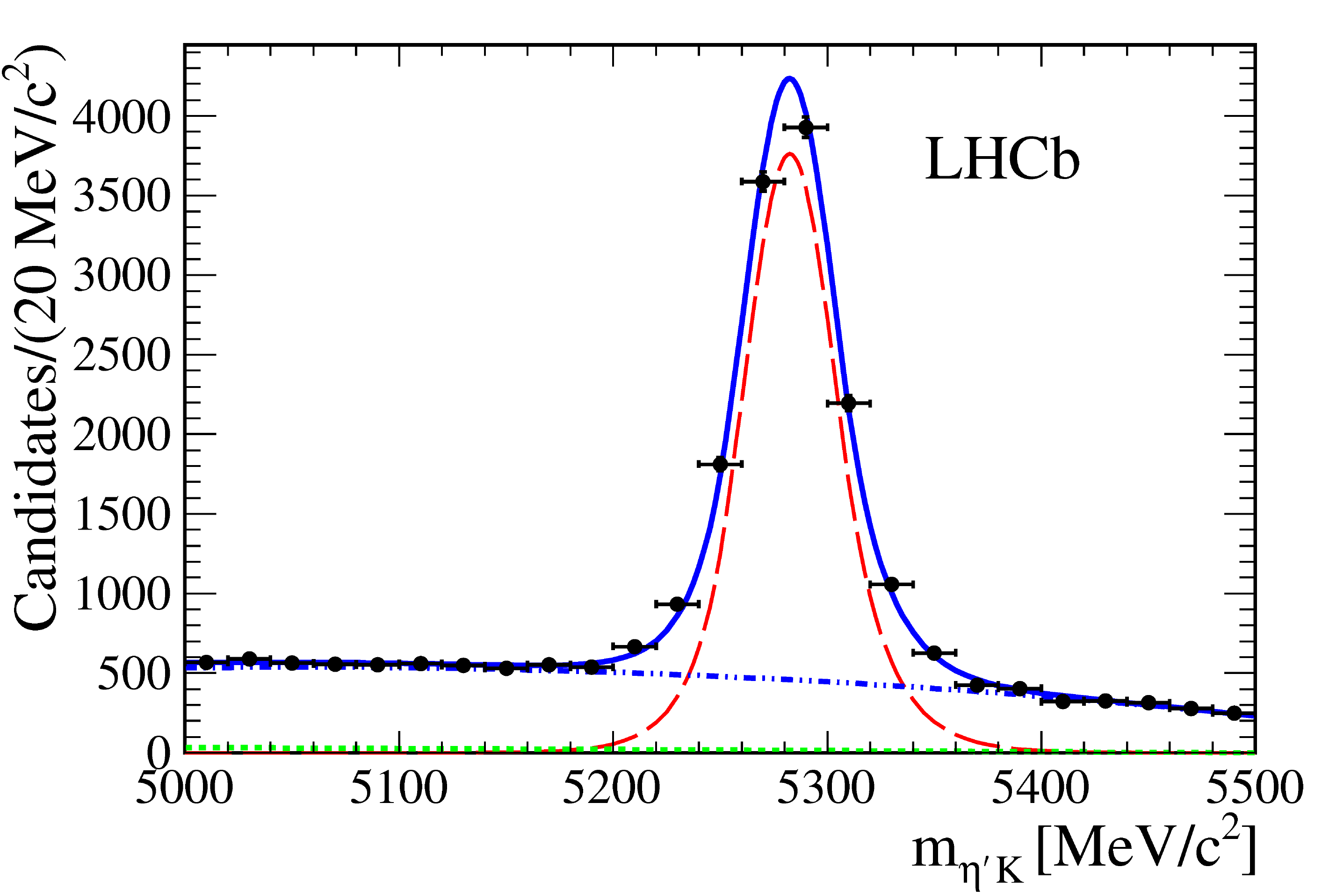}
\caption{Distributions of the (top left) $\pi^+\pi^-\gamma$ and (top right)  $\etapr K^+K^-$ masses of the selected
\decay{\Bs}{\etapr\phiz} candidates, as well as of the (bottom left) $\pi^+\pi^-\gamma$  and (bottom right)  $\etapr \Kp$ masses of the selected
\decay{\Bp}{\etapr\Kp} candidates. 
The solid blue curves represent the result of the simultaneous two-dimensional fit described in the text, with the following components:
\decay{\Bs}{\etapr\phiz} and \decay{\Bp}{\etapr\Kp} signals (red dashed), combinatorial backgrounds (blue  dot-dot-dashed), combinatorial backgrounds with real $\etapr$ (green dotted), and \decay{\Bs}{\phiz\phiz} background (black dot-dashed). } 
\label{fig:fitplots}
\end{figure}
 
Figure~\ref{fig:fitplots} shows the mass distributions observed in data with the projections of the fit results overlaid.
No \decay{\Bs}{\etapr\phiz} signal is observed.
The fitted yields are $-3.2^{+5.0}_{-3.8}$ for the \decay{\Bs}{\etapr\phiz} signal, $105 \pm 29$ for the 
\decay{\Bs}{\phiz\phiz} physics background (consistent with expectation), and $11\,081 \pm127$ 
for the \decay{\Bp}{\etapr\Kp} normalisation mode. 
The measured $\B$ and $\etapr$ mass resolutions are \mbox{$21.8\pm 0.3\mevcc$} and \mbox{$12.6\pm 0.2\mevcc$}, respectively.
The ratio of yields is fitted to be
\mbox{$N(\decay{\Bs}{\etapr\phiz})/N(\decay{\Bp}{\etapr\Kp}) = (-2.9^{+4.5}_{-3.5})\times 10^{-4}$}. 

Sets of pseudoexperiments are used to evaluate possible fit biases.
Fits on samples generated from the probability density function (PDF) with parameters obtained from the data are found to be unbiased.
The procedure is then repeated using simulated \decay{\Bs}{\phiz\phiz} events instead of
generating the corresponding background component from the PDF. 
Biases of $-1.3\pm0.3$ on the signal yield and of $(-1.16\pm 0.33)\times10^{-4}$ on the ratio of yields are observed.
The results obtained with data are corrected for these biases and systematic uncertainties computed
as the quadratic sum of the statistical uncertainty on the bias and half of the bias value are assigned.

Additional systematic uncertainties affect the signal yield and the yield ratio.
The mass fit is repeated with different combinatorial background PDFs:
linear functions are replaced with exponential functions, and the parabolic function is replaced
with a third-order polynomial.
The quadratic sum of the differences between the values obtained in these alternative fits
and the nominal result is assigned as a systematic uncertainty. 
The limited size of the simulated  \decay{\Bs}{\phiz\phiz} sample leads to an uncertainty on the determination
of the nonparametric PDF for the physics background, which is propagated as a systematic uncertainty.
The effect of fixing some of the model parameters in the fit is studied by performing a large number of fits on the data, with the fixed
parameters sampled randomly from Gaussian distributions centred on the nominal values and with
widths and correlations as determined in simulated events. The standard deviation of the distribution of the
results is assigned as a systematic uncertainty.

The  systematic uncertainties on the $\Bs \to \etapr \phiz$ yield and the yield ratio are summarised in Table~\ref{tab:Systematicsfit}.
The final results from the mass fit, including all corrections, are 
\begin{eqnarray}
N(\Bs \to \etapr \phiz) &=& -1.9^{+5.0}_{-3.8} \pm 1.1 \,, \\
\frac{N(\Bs \to \etapr \phiz)}{N(B^+ \to \etapr K^+)} &=& (-1.7^{+4.5}_{-3.5} \pm 1.0)\times 10^{-4} \,,
\label{eq:yield_ratio}
\end{eqnarray}
where the first (second) quoted uncertainty is statistical (systematic). 
Bayesian upper limits $x_{\rm U}$ are determined
assuming a uniform prior in the observable $x$ (yield, yield ratio, or $\cal B$)
as $\int_0^{x_{\rm U}} {\cal L}(x) dx /  \int_0^{\infty} {\cal L}(x) dx = \alpha$,
where ${\cal L}(x)$ is the likelihood function convolved
with the systematic uncertainties, and $\alpha$ is the confidence level (CL).
The obtained upper limits are 
\begin{equation}
N(\decay{\Bs}{\etapr\phiz})< 8.9~(10.9) \quad \mbox{at  90\% (95\%) CL} \nonumber
\end{equation}
and
\begin{equation}
\frac{N(\decay{\Bs}{\etapr \phiz})}{N(B^+ \to \etapr K^+)}< 8.0~(9.9)\times 10^{-4} \quad \mbox{at  90\% (95\%) CL}\,. \nonumber
\end{equation}

\begin{table}[!t]
\caption{Systematic uncertainties $\sigma_N$ and $\sigma_R$ on the fitted yield
$N(\decay{\Bs}{\etapr\phiz})$ and on the yield ratio 
$R= N(\decay{\Bs}{\etapr\phiz})/N(\decay{\Bu}{\etapr K^+})$, respectively.
The last line gives the quadratic sum of the individual uncertainties.}
\begin{center}
\begin{tabular}{lcc}
\hline
Source & $\sigma_N$ (events) &  $\sigma_R$ ($10^{-4}$)   \\  
\hline
Fit bias   & 0.7 & 0.7\\ 
Combinatorial background modelling & 0.6 & 0.6\\ 
$\decay{\Bs}{\phiz\phiz}$ background modelling  & 0.4 & 0.3\\ 
Fixed parameters in the fit & 0.3 & 0.3\\ 
\hline
Total   & 1.1 & 1.0\\ 
\hline
\end{tabular}\end{center}
\label{tab:Systematicsfit}
\end{table}

\section{Result and conclusion}

A search has been performed for the \decay{\Bs}{\etapr\phiz} decay. 
No signal is found. 
The branching fraction 
$\BR(\decay{\Bs}{\etapr \phiz}) = (-0.18^{+0.47}_{-0.36}({\rm stat}) \pm  0.10 ({\rm syst}))\times 10^{-6}$ is computed from
Eqs.~(\ref{eq:master}), (\ref{eq:eff_ratio}) and (\ref{eq:yield_ratio}) using the known value of 
$\BR(\decay{\Bp}{\etapr \Kp})$~\cite{PDG2016} and the LHCb measurement of $f_s/f_d$~\cite{fsfd}, which leads to
\begin{equation}
{\cal B}(\decay{\Bs}{\etapr\phiz})< 0.82\,(1.01)\times 10^{-6} \quad \mbox{at  90\% (95\%) CL}\, \nonumber
\end{equation}
using the likelihood integration method described above.
This is the first upper limit set on the \decay{\Bs}{\etapr\phiz} branching fraction.

This result favours the lower end of the range of predictions for this branching fraction, pointing to form factors consistent with the
light-cone sum-rule calculation used in Ref.~\cite{Cheng:2009mu}, or with the hypotheses used
in Refs.~\cite{Beneke:2003zv,Ali:2007ff}.
Although large theoretical uncertainties make most predictions compatible with the result of this analysis, the central values
of the predictions in Refs.~\cite{Chen:2007qm,Wang:2008rk,Cheng:2014rfa, Zhou:2016jkv}
are significantly larger than the upper limit. These discrepancies should help in constraining
the theoretical models used in the prediction of branching fractions and \CP asymmetries for $B$-meson hadronic charmless decays.

\section*{Acknowledgements}

 
\noindent We express our gratitude to our colleagues in the CERN
accelerator departments for the excellent performance of the LHC. We
thank the technical and administrative staff at the LHCb
institutes. We acknowledge support from CERN and from the national
agencies: CAPES, CNPq, FAPERJ and FINEP (Brazil); NSFC (China);
CNRS/IN2P3 (France); BMBF, DFG and MPG (Germany); INFN (Italy); 
FOM and NWO (The Netherlands); MNiSW and NCN (Poland); MEN/IFA (Romania); 
MinES and FASO (Russia); MinECo (Spain); SNSF and SER (Switzerland); 
NASU (Ukraine); STFC (United Kingdom); NSF (USA).
We acknowledge the computing resources that are provided by CERN, IN2P3 (France), KIT and DESY (Germany), INFN (Italy), SURF (The Netherlands), PIC (Spain), GridPP (United Kingdom), RRCKI and Yandex LLC (Russia), CSCS (Switzerland), IFIN-HH (Romania), CBPF (Brazil), PL-GRID (Poland) and OSC (USA). We are indebted to the communities behind the multiple open 
source software packages on which we depend.
Individual groups or members have received support from AvH Foundation (Germany),
EPLANET, Marie Sk\l{}odowska-Curie Actions and ERC (European Union), 
Conseil G\'{e}n\'{e}ral de Haute-Savoie, Labex ENIGMASS and OCEVU, 
R\'{e}gion Auvergne (France), RFBR and Yandex LLC (Russia), GVA, XuntaGal and GENCAT (Spain), Herchel Smith Fund, The Royal Society, Royal Commission for the Exhibition of 1851 and the Leverhulme Trust (United Kingdom).

\addcontentsline{toc}{section}{References}
\setboolean{inbibliography}{true}
\bibliographystyle{LHCb}
\bibliography{main,LHCb-PAPER,LHCb-CONF,LHCb-DP,LHCb-TDR}

\newpage


 \newpage
\centerline{\large\bf LHCb collaboration}
\begin{flushleft}
\small
R.~Aaij$^{40}$,
B.~Adeva$^{39}$,
M.~Adinolfi$^{48}$,
Z.~Ajaltouni$^{5}$,
S.~Akar$^{59}$,
J.~Albrecht$^{10}$,
F.~Alessio$^{40}$,
M.~Alexander$^{53}$,
S.~Ali$^{43}$,
G.~Alkhazov$^{31}$
% ...

{\footnotesize \it
% ...
$ ^{5}$Clermont Universit{\'e}, Universit{\'e} Blaise Pascal, CNRS/IN2P3, LPC, Clermont-Ferrand, France\\
% ...
$ ^{10}$Fakult{\"a}t Physik, Technische Universit{\"a}t Dortmund, Dortmund, Germany\\
% ...
$ ^{31}$Petersburg Nuclear Physics Institute (PNPI), Gatchina, Russia\\
% ...
$ ^{39}$Universidad de Santiago de Compostela, Santiago de Compostela, Spain\\
$ ^{40}$European Organization for Nuclear Research (CERN), Geneva, Switzerland\\
% ...
$ ^{43}$Nikhef National Institute for Subatomic Physics, Amsterdam, The Netherlands\\
% ...
$ ^{48}$H.H. Wills Physics Laboratory, University of Bristol, Bristol, United Kingdom\\
% ...
$ ^{53}$School of Physics and Astronomy, University of Glasgow, Glasgow, United Kingdom\\
% ...
$ ^{59}$University of Cincinnati, Cincinnati, OH, United States\\
% ...
}
\end{flushleft}

\end{document}